# Solutions to some Molecular Potentials in *D*-Dimensions: Asymptotic Iteration Method.

D. Agboola [a]

*Department of Pure and Applied Mathematics,*

*Ladoke Akintola University of Technology,*

*Oyo State, Nigeria. P.M.B. 4000.*

**Abstract.**

We give the study of the bound-state solutions of some molecular vibration potentials by solving the *D*-dimensional Schrödinger equation using the asymptotic iteration method (AIM). The eigenvalue values and the corresponding eigenfunctions are also obtained using the AIM. It was found that the asymptotic iteration method gives the eigenvalues directly by some transformation of the radial Schrödinger equation; likewise, the asymptotic iteration method yields exact analytical solutions for exactly solvable problems and provides the closed-forms for the energy eigenvalues as well as the corresponding eigenfunctions.

## 1. Introduction

Since its introduction [1], the asymptotic iterative method (AIM) has widely been applied in solving many eigenvalue problems in both the relativistic [2-4] and non-relativistic [5-18] quantum. The AIM has been used in solving the Schrödinger equation for hydrogen-like atom[11] exponential-type potentials such as the Hulthén potential [13], the Morse Potential [10,14, 16], and some singular potentials like the generalized spiked harmonic potential [6]. Recently, an iterative treatment of the relativistic Dirac equation with the Coloumbic potential was presented using the AIM [3].

However, over the decades, a thorough research has been carried out on some molecular vibration potentials. For instance, some quantum mechanical properties of the Kratzer-Fues potential have been presented in *N*-dimensions by Oyewumi [22]. Likewise, the exact solutions of the Mie-type Potentials in *D*-dimension have been discussed by Ikhdair and Server [23]. Recently, Agboola *et al* [28] gave an *N*-dimensional study of the Pseudoharmonic potential.

Therefore the aim of this letter is to give the bound-state solutions of the Schrödinger equation with some molecular potentials in *D*-dimensions using the AIM. The paper is organized as follows: Section 2 gives a brief description of the AIM, while in the following three sections we obtain the eigenvalues and eigenfunctions of some molecular potentials using the AIM. Numerical results are given in the tables. Finally, we conclude by discussing the various result obtained.

## 2. The Asymptotic Iterative Method

In this section, we give a brief description of the AIM; details of the method can be obtained in Refs [1, 18]. Suppose we wish to solve the homogenous linear second-order differential equation

$$y'' = f_0(x) y' + g_0(x) y, \quad f_0(x) \neq 0 \tag{1}$$

Where $f_o(x)$ and $g_o(x)$ have sufficiently many continuous derivatives and defined in some interval which are not necessarily bounded. Due to the symmetric structure of the right hand side of Equation (1), we can have the (n+1)th and (n+2)th derivatives of (1) as follows:

$$y^{(n+1)} = f_{n-1}(x)y' + g_{n-1}(x)y \qquad (2)$$

and

$$y^{(n+2)} = f_n(x)y' + g_n(x)y \qquad (3)$$

with the relation

$$f_n = f'_{n-1} + g_{n-1} + f_o f_{n-1} \text{ and } g_n = g'_{n-1} + g_o f_{n-1} \qquad (4)$$

From the ratio of the (n+1)th and (n+2)th derivatives, we have

$$\frac{d}{dx} \log_e y^{(n+1)} = \frac{y^{(n+2)}}{y^{(n+1)}} = \frac{f_n\left(y' + \frac{g_n}{f_n}y\right)}{f_{n-1}\left(y' + \frac{g_{n-1}}{f_{n-1}}y\right)} \qquad (5)$$

For a sufficiently large $n$, we can have the following asymptotic expression

$$\frac{g_n}{f_n} = \frac{g_{n-1}}{f_{n-1}} = \alpha(x) \qquad (6)$$

with the termination condition given as

$$\Delta_k(x) = \begin{vmatrix} g_k & f_k \\ g_{k-1} & f_{k-1} \end{vmatrix} = g_k f_{k-1} - f_k g_{k-1} = 0 \quad k = 1,2,3,\ldots \qquad (7)$$

We also note that the energy eigenvalues are obtained from the roots of the equation (7) if the problem is exactly solvable. However, for a specific $n$ principal quantum number, we choose a suitable $x_o$ point, determined generally as the maximum value of the asymptotic wave function or the minimum value of the potential [3, 18, 19, 20], and the approximate energy eigenvalues are obtained from the roots of this equation for sufficiently great values of k with iteration.

Using Equation (6), Equation (5) reduces to

$$\frac{d}{dx}\log_e y^{(n+1)} = \frac{f_n}{f_{n-1}} \tag{8}$$

which yields

$$y^{(n+1)}(x) = C_1 \exp\left(\int_x \frac{f_n(t)}{f_{n-1}(t)} dt\right) = C_1 \exp f_{n-1}\left(\int_x (\alpha + f_o) dt\right) \tag{9}$$

Note that we have used the relations (4) and (6) in obtaining the right hand side of Equation (9) and $C_1$ is the integration constant. Substituting Equation (9) into (2), we have the first order differential equation

$$y' + \alpha y = C_1 \exp\left(\int_x (\alpha + f_o) dt\right) \tag{10}$$

Solving Equation (10), we have the general solution to Equation (1) as follows:

$$y(x) = \exp\left(-\int_x \alpha dt\right)\left[C_2 + C_1 \int_x \exp\left(\int_x (f_o(\tau) + 2\alpha(\tau)) d\tau\right) dt\right] \tag{11}$$

### 3. The Harmonic Oscillator potential in *D*-Dimensions

First, we start by studying the harmonic oscillator potential in *D*-dimensions [26]. The Schrödinger equation for the oscillator can be written as:

$$H\Psi_{n,\ell,m}(r,\Omega) = E\Psi_{n,\ell,m}(r,\Omega) \tag{12}$$

where *E* is the energy eigenvalue, $\Psi_{n,\ell,m}(r,\Omega)$ is the wave function and *H* is the Hamiltonian given as:

$$H = \frac{\hbar^2}{2\mu}\left[r^{1-D}\frac{\partial}{\partial r}\left(r^{D-1}\frac{\partial}{\partial r}\right) + \frac{\Lambda_D^2}{r^2}\right] + \frac{\mu\omega^2 r^2}{2} \tag{13}$$

Inserting equations (12) and (11) and separating the variables as follows

$$\Psi_{n,\ell,m}(r,\Omega) = R_{n,\ell}(r) Y_\ell^m(\Omega), \tag{14}$$

Equation (11) reduces to two separate equations namely:

$$R''(r) + \frac{D-1}{r} R'(r) - \frac{\ell(\ell+D-2)}{r^2} R(r) + \frac{2\mu}{\hbar^2}\left[E - \frac{\mu\omega^2 r^2}{2}\right] R(r) = 0 \tag{15}$$

and

$$\Lambda_D^2(\Omega) Y_\ell^m(\Omega) + \beta Y_\ell^m(\Omega) = 0 \tag{16}$$

where $\beta$ is the separation constant given as

$$\beta = \ell(\ell+D-2), \quad \ell = 0,1,2,\ldots \tag{17}$$

and $\ell$ is the angular momentum quantum number.

However, with the behavior of the radial function at zero and infinity, we can have the asymptotic solution to Equation (13) as follows.

$$R_{n,\ell}(r) = r^\ell \exp\left(-\frac{\mu\omega r^2}{2\hbar}\right) U_{n,\ell}(r) \tag{18}$$

With this, Equation (14) becomes

$$U''_{n,\ell}(r) = \left[2\gamma r - \frac{2\ell+D-1}{r}\right] U'_{n,\ell}(r) + \left[\gamma(2\ell+D) + \varepsilon^2\right] U_{n,\ell}(r) \tag{19}$$

where $\gamma = \frac{\mu\omega}{\hbar}$ and $-\varepsilon^2 = \frac{2\mu E}{\hbar^2}$.

We now apply the AIM in solving Equation (19). Comparing Equations (1) and (19), we have

$$f_o(r) = 2\gamma r - \frac{2\ell + D - 1}{r} \quad \text{and} \quad g_o(r) = \gamma(2\ell + D) + \varepsilon^2 \tag{20}$$

with the use for the relation (4), we have the following derivatives:

$$f_1(r) = 4\gamma^2 r^2 + \frac{(2\ell + D - 1)(2\ell + D)}{r^2} + 4\gamma(\gamma r + 2\ell + D - 1) + \gamma(2\ell + D) + \varepsilon^2$$

$$g_1(r) = \left[2\gamma r + \frac{(2\ell + D - 1)}{r^2}\right]\left[\gamma(2\ell + D) + \varepsilon^2\right]$$

$$f_2(r) = 8\gamma^2 r - \frac{2(2\ell + D - 1)(2\ell + D)}{r^3} + \left[2\gamma r + \frac{(2\ell + D - 1)}{r^2}\right]\left[\gamma(2\ell + D) + \varepsilon^2\right] + \left[2\gamma r + \frac{(2\ell + D - 1)}{r}\right]f_1$$

$$g_2(r) = \left(\gamma(2\ell + D) + \varepsilon^2\right)\left[2\gamma + \frac{(2\ell + D - 1)}{r^3}\right] + \left(\gamma(2\ell + D) + \varepsilon^2\right)f_1$$

…… etc. (21)

Employing the terminating condition (7), we arrive at the following eigenvalue expression:

for k = 1: $g_1 f_0 - g_0 f_1 = 0 \quad \Rightarrow \quad -\varepsilon_0^2 = \gamma(2\ell + D)$

for k = 2: $g_2 f_1 - g_1 f_2 = 0 \quad \Rightarrow \quad -\varepsilon_1^2 = \gamma(2\ell + D + 2)$

for k = 3: $g_3 f_2 - g_2 f_3 = 0 \quad \Rightarrow \quad -\varepsilon_2^2 = \gamma(2\ell + D + 4)$

….. etc. (22)

Generalizing the above expressions and using the identities in Equation (19), we have the energy eigenvalues as follows:

$$E_{n,\ell} = \hbar\omega\left(\ell + n + \frac{D}{2}\right) \quad n, \ell = 0,1,2,3,\ldots \tag{23}$$

With $\ell = 0$ and $D = 1$, the energy values becomes

$$E_n = \hbar\omega\left(n + \tfrac{1}{2}\right) \qquad n = 0,1,2,3,\ldots \qquad (24)$$

which is in agreement with the values obtained for *s*-state [28].

Furthermore, we now want to obtain the eigenfunctions using the AIM. Generally speaking, to obtain the eigenfunctions using AIM, the differential equation we wish to solve is of the form [19]:

$$y''(x) = 2\left(\frac{ax^{N+1}}{1-bx^{N+2}} - \frac{t+1}{x}\right)y'(x) - \frac{wx^N}{1-bx^{N+2}}y(x) \qquad 0 < x < \infty \qquad (25)$$

where $a$ and $b$ are constants and $w$ can be determined from the condition (6) for $k = 0, 1, 2, 3, \ldots$ and $N = -1,0,1,2,3,\ldots$ the general solution of (25) is given as

$$y_n(x) = (-1)^n C_1 (N+2)^n (\sigma)_n \,{}_2F_1\left(-n, \rho+n; \sigma; bx^{N+2}\right) \qquad (26)$$

where

$$(\sigma)_n = \frac{\Gamma(\sigma+n)}{\Gamma(\sigma)} \qquad \sigma = \frac{2t+N+3}{N+2} \quad \text{and} \quad \rho = \frac{(2t+1)b+2a}{(N+2)b} \qquad (27)$$

Comparing Equations (19) and (25) we have $N = 0$, $a = \gamma$, $b = 0$ and $t = (2\ell + D - 3)/2$. Therefore, we find $\rho = \frac{2\ell + D - 2}{2} + \frac{\gamma}{b}$ and $\sigma = \frac{2\ell + D}{2}$. So, the solutions to Equations (19) are given as follows:

$$U_{n\ell}(r) = (-1)^n 2^n \frac{\Gamma\left(n + \frac{2\ell+D}{2}\right)}{\Gamma\left(\frac{2\ell+D}{2}\right)} \,{}_1F_1\left(-n; \frac{2\ell+D}{2}; \frac{\mu\omega r^2}{\hbar}\right) \qquad (28)$$

Note that we have use the following limit expression of the hypergeometric function:

$$\lim_{b \to 0} {}_2F_1\left(-n, \frac{\gamma}{b} + a; c; zb\right) = {}_1F_1(-n; c; z) \qquad (29)$$

Equations (18) and (28) give the eigenfunctions as

$$R_{n,\ell}(r) = C_{n,\ell}(-1)^n 2^n \frac{\Gamma\left(n + \frac{2\ell+D}{2}\right)}{\Gamma\left(\frac{2\ell+D}{2}\right)} r^\ell \exp\left(-\frac{\mu\omega r^2}{2\hbar}\right){}_1F_1\left(-n; \frac{2\ell+D}{2}; \frac{\mu\omega r^2}{\hbar}\right) \tag{30}$$

Where $C_{n,\ell}$ is the normalization constant.

## 4. The Pseudoharmonic Potential in *D*-Dimensions

In this section, we study the bound state solution of the Pseudoharmonic potential given as [29-32]:

$$V(r) = \frac{1}{8}\kappa r_e^2 \left(\frac{r}{r_e} - \frac{r_e}{r}\right)^2 \tag{31}$$

where $r_e$ is the equilibrium bond length and $\kappa$ is the force constant. The hyperradial part of the Schrödinger equation with the Pseudoharmonic oscillator in *D*-dimensions can be written as:

$$R''_{n,\ell}(r) + \frac{D-1}{r} R'_{n,\ell}(r) + \left[-\varepsilon^2 + \frac{\mu\kappa r_e}{2\hbar^2} - \frac{\mu\kappa r^2}{4\hbar^2} - \frac{v(v+D-2)}{r^2}\right]R_{n,\ell}(r) = 0 \tag{32}$$

where $\mu$ is the reduced mass, $-\varepsilon^2 = \frac{2\mu E}{\hbar^2}$ and $v(v+D-2) = \ell(\ell+D-2) + \frac{\mu\kappa r_e^4}{4\hbar^2}$.

According to the asymptotic behaviors of the wave function as $r \to 0$ and $r \to \infty$, one can express the solution as

$$R_{n,\ell}(r) = r^v \exp\left(-\sqrt{\frac{\mu\kappa}{16\hbar^2}} r^2\right) U_{n\ell}(r) \tag{33}$$

With the use of (33), Equation (32) becomes

$$U''_{n\ell}(r) = \left[\sqrt{\frac{\mu\kappa}{\hbar^2}} r - \frac{(2v+D-1)}{r}\right]U'_{n\ell}(r) + \left[\frac{1}{2}(2v+D)\sqrt{\frac{\mu\kappa}{\hbar^2}} - \frac{\mu\kappa r_0^2}{2\hbar^2} + \varepsilon^2\right]U_{n\ell}(r) \tag{34}$$

We now solve Equation (34) using the AIM. Comparing Equations (1) and (34), and using the recursion relation (4), we obtain the following:

$$f_0(r) = \sqrt{\frac{\mu\kappa}{\hbar^2}}\,r - \frac{(2v+D-1)}{r}$$

$$g_0(r) = \frac{1}{2}(2v+D)\sqrt{\frac{\mu\kappa}{\hbar^2}} - \frac{\mu\kappa r_0^2}{2\hbar^2} + \varepsilon^2$$

$$f_1(r) = \frac{\mu\kappa}{\hbar^2}r^2 + \frac{(2v+D-1)(2v+D)}{r^2} - \frac{3}{2}\sqrt{\frac{\mu\kappa}{\hbar^2}}(2v+D-2) - \frac{\mu\kappa r_0^2}{2\hbar^2} + \varepsilon^2$$

$$g_1(r) = \left[\sqrt{\frac{\mu\kappa}{\hbar^2}}\,r - \frac{(2v+D-1)}{r}\right]\left[\frac{1}{2}(2v+D)\sqrt{\frac{\mu\kappa}{\hbar^2}} - \frac{\mu\kappa r_0^2}{2\hbar^2} + \varepsilon^2\right]$$

$$f_2(r) = \frac{2\mu\kappa}{\hbar^2}r - \frac{(2v+D-1)(2v+D)}{r^3} +$$
$$\left[\sqrt{\frac{\mu\kappa}{\hbar^2}}\,r - \frac{(2v+D-1)}{r}\right]\left[\frac{\mu\kappa}{\hbar^2}r^2 + \frac{(2v+D-1)(2v+D)}{r^2} - \sqrt{\frac{\mu\kappa}{\hbar^2}}(2v+D-3) - \frac{\mu\kappa r_0^2}{\hbar^2} + 2\varepsilon^2\right]$$

$$g_2(r) = \left[\frac{\mu\kappa}{\hbar^2}r^2 + \frac{(2v+D-1)(2v+D+1)}{r^2} - \frac{3}{2}\sqrt{\frac{\mu\kappa}{\hbar^2}}(2v+D-2) + \sqrt{\frac{\mu\kappa}{\hbar^2}} - \frac{\mu\kappa_0^2}{2\hbar^2} + \varepsilon^2\right]\left[\frac{1}{2}(2v+D)\sqrt{\frac{\mu\kappa}{\hbar^2}} - \frac{\mu\kappa_0^2}{2\hbar^2} + \varepsilon^2\right]$$

……etc. (35)

In similar fashion with the previous section, using the termination condition (7), gives the following expressions:

$$\text{for k} = 1:\; -\varepsilon_0^2 = \frac{1}{2}(2v+D)\sqrt{\frac{\mu\kappa}{\hbar^2}} - \frac{\mu\kappa r_0^2}{2\hbar^2}$$

$$\text{for k} = 2:\; -\varepsilon_1^2 = \frac{1}{2}(4+2v+D)\sqrt{\frac{\mu\kappa}{\hbar^2}} - \frac{\mu\kappa r_0^2}{2\hbar^2}$$

$$\text{for k} = 3:\; -\varepsilon_2^2 = \frac{1}{2}(8+2v+D)\sqrt{\frac{\mu\kappa}{\hbar^2}} - \frac{\mu\kappa r_0^2}{2\hbar^2}$$

......*etc.* (36)

Generalizing the above expression and using the identities in Equation (32), we have the energy eigenvalues for the Pseudoharmonic potential

$$E_{n,\ell} = \frac{1}{4}(4n + 2v + D)\sqrt{\frac{\kappa\hbar^2}{\mu}} - \frac{\kappa r_0^2}{4} \qquad n = 0,1,2,3,... \tag{37}$$

and

$$v = \frac{1}{2}\left[\frac{(2-D)}{2} + \sqrt{(2\ell + D - 2)^2 + \frac{\mu\kappa r_0^4}{\hbar^2}}\right] \tag{38}$$

By comparing Equations (25) and (34), we have the relations: $N = 0$, $b = 0$, $a = \frac{1}{2}\sqrt{\frac{\mu\kappa}{\hbar^2}}$ and $t = (2v + D - 3)/2$. And so, with the use of Equations (26), (27) and (29), we have

$$U_{n\ell}(r) = (-1)^n 2^n \frac{\Gamma\left(n + \frac{2v+D}{2}\right)}{\Gamma\left(\frac{2v+D}{2}\right)} {}_1F_1\left(-n; \frac{2v+D}{2}; \sqrt{\frac{\mu\kappa}{4\hbar^2}}r^2\right) \tag{39}$$

Equations (33) with (39) give the unnormalized eigenfunctions

$$R_{n,\ell}(r) = C_{n,\ell}(-1)^n 2^n \frac{\Gamma\left(n + \frac{2v+D}{2}\right)}{\Gamma\left(\frac{2v+D}{2}\right)} r^v \exp\left(-\sqrt{\frac{\mu\kappa}{16\hbar^2}}r^2\right) {}_1F_1\left(-n; \frac{2v+D}{2}; \sqrt{\frac{\mu\kappa}{4\hbar^2}}r^2\right) \tag{40}$$

where $C_{n,\ell}$ is the normalization constant.

## 5. The Kratzer-Fues Potential in *D*-Dimensions.

We now turn our attention to the Mie-type potential-the Kratzer-Fues potential [33]. Although the bound state of this potential has been discuss in *D*-dimensions using the polynomial method [22, 23] and the Nikiforov-Uvarov method [24]. However, in this section, we wish to obtain the eigenvalues and eigenfunctions of the potential using the AIM. Following the notations in [22], we write the Kratzer-Fues potential as:

$$V(r) = -\frac{A}{r} + \frac{B}{r^2} \tag{41}$$

Where $A = 2D_e r_0$ and $B = D_e r_0^2$, $D_e$ is the interaction energy between two atoms in a molecular system at distance $r = r_0$. The eigenvalue equation for the potential in D-dimension is given as:

$$R''_{nl}(r) + \frac{D-1}{r} R'_{nl}(r) + \left[ -\varepsilon^2 + \frac{\alpha}{r} - \frac{v(v+D-2)}{r^2} \right] R_{nl}(r) = 0 \tag{42}$$

where we have define $-\varepsilon^2 = \frac{2\mu E}{\hbar^2}$, $\alpha = \frac{2\mu A}{\hbar^2}$ and $v(v+D-2) = \ell(\ell+D-2) + \frac{2\mu B}{\hbar^2}$. If one defines a new variable $z = 2\varepsilon r$ and then assume the solution is of the form

$$R_{nl}(z) = z^v \exp(-z/2) U_{nl}(z) \quad , \tag{43}$$

Equation (42) becomes

$$U''_{nl}(z) = \left[ 1 + \frac{(2v+D-1)}{z} \right] U'_{nl}(z) + \left[ \frac{2v+D-1}{2} - \frac{\alpha}{2\varepsilon} \right] \frac{U_{nl}(z)}{z} \tag{44}$$

Comparing Equations (1) and (44), and using the relation (4) we have the following:

$$f_0 = \left[ 1 + \frac{(2v+D-1)}{z} \right]$$

$$g_0 = \frac{1}{2z}\left[(2v+D-1) + \alpha/\varepsilon\right]$$

$$f_1 = \frac{(2v+D-1)(2v+D-2)}{z^2} + \frac{1}{2z}\left[5(2v+D-1) + \alpha/\varepsilon\right] + 1$$

$$g_1 = \frac{(2v+D-2)}{2z^2}\left[(2v+D-1) - \alpha/\varepsilon\right] + \frac{1}{2z}\left[(2v+D-1) + \alpha/\varepsilon\right]$$

……etc. (45)

The termination condition (7) therefore yields for k = 1: $\varepsilon_0 = \dfrac{\alpha}{2v + D - 1}$

for k = 2: $\varepsilon_1 = \dfrac{\alpha}{2 + 2v + D - 1}$

for k = 3: $\varepsilon_2 = \dfrac{\alpha}{4 + 2v + D - 1}$

………etc. (46)

Generalizing Equation (46) and with the use of $-\varepsilon^2 = \dfrac{2\mu E}{\hbar^2}$ and $\alpha = \dfrac{2\mu A}{\hbar^2}$, we get the energy eigenvalue of the Kratzer-Fues potential as follows

$$E_{n\ell} = \dfrac{-2\mu A^2}{\hbar^2 [2n + 2v + D - 1]^2} \qquad n = 0,1,2,3,... \qquad (47)$$

where

$$v = \dfrac{1}{2}\left[(2 - D) + \sqrt{(2\ell + D - 2)^2 + \dfrac{8\mu B}{\hbar^2}}\right] \qquad (48)$$

Following the argument presented in the previous sections, if we comparing Equations (25) and (44), we can define $N = -1$, $a = \dfrac{1}{2}$, $b = 0$ and $t = \dfrac{2v + D - 3}{2}$. Consequently, with the use of Equations (26), (27) and (29) we have the solution to (44) as

$$U_{n\ell}(z) = (-1)^n \dfrac{\Gamma(n + 2v + D)}{\Gamma(2v + D)} {}_1F_1(-n; 2v + D; z) \qquad (49)$$

Finally, Equation (43) with (49) gives the wavefunction of the Kratzer-Fues potential

$$R_{n\ell}(r) = C_{n\ell}(-1)^n \dfrac{\Gamma(n + 2v + D)}{\Gamma(2v + D)} (2\varepsilon r)^v \exp(-\varepsilon r) {}_1F_1(-n; 2v + D; 2\varepsilon r) \qquad (50)$$

$C_{n\ell}$ is the normalization constant.

## 6. Conclusion

In this paper, some potentials describing the vibrations in a molecular system has been discussed in *D*-dimensions within the work frame of the asymptotic iteration method. The energy eigenvalues were found to be in good agreement with those obtained using the direct integration method and the Nikiforov-Uvarov method. The eigenfunctions were also obtained in terms of the hypergeometric function using the AIM.

It is pertinent to note that the asymptotic iteration method gives the eigenvalues directly by transforming the radial Schrödinger equation into a form of $y'' = f_0(x)y' + g_0(x)y$.

In comparison, the asymptotic iteration method yields exact analytical solutions for exactly solvable problems and provides the closed-forms for the energy eigenvalues as well as the corresponding eigenfunctions. However, where there is no such a solution, the energy eigenvalues are obtained by using an iterative approach [34-37].

Moreover, AIM puts no constraint on the potential parameter values involved and it is easy to implement. The results are sufficiently accurate for practical purposes.